%% file: bottle1510-ps.tex
\documentclass[10pt]{article}
\usepackage{amsmath,amsfonts,amssymb,a4,color,epsfig,graphics}
\usepackage[latin1]{inputenc}
\usepackage{verbatim}
\def\pa{\partial}        
\def\ess{essentially self-adjoint}

\def\ra{\rightarrow}

\def\bR {{\mathbb{R}}}
\def\bN {{\mathbb{N}}}
\def\bC {{\mathbb{C}}}

\def\bZ {{\mathbb{Z}}}

\def\bS {{\mathbb{S}}}

\makeatletter
\@addtoreset{equation}{section}

\makeatother

\newtheorem{theorem}{Theorem}[section]
\newtheorem{lemma}[theorem]{Lemma}
\newtheorem{proposition}[theorem]{Proposition}
\newtheorem{definition}[theorem]{Definition}

\newtheorem{remark}[theorem]{Remark}

\begin{document}

\bibliographystyle{plain}

\title{ Confining quantum particles\\ with a purely magnetic field}

\author{Yves Colin de Verdi\`ere\footnote{Institut Fourier,
 Unit{\'e} mixte
 de recherche CNRS-UJF 5582,
 BP 74, 38402-Saint Martin d'H\`eres Cedex (France),
yves.colin-de-verdiere@ujf-grenoble.fr}~ \&~
 Fran\c{c}oise Truc\footnote{Institut Fourier, 
francoise.truc@ujf-grenoble.fr}}

\maketitle

\begin{abstract}
We consider  a 
 Schr\"odinger operator with a magnetic field (and no electric field)
 on a domain  in the Euclidean space  with a compact boundary.
 We give sufficient conditions on the behaviour  of the magnetic
 field  near the boundary which
guarantees essential self-adjointness of this operator.
From the physical point of view, it means that the quantum 
particle is confined in the domain by the magnetic field.
We construct examples in the case where the boundary
is smooth as well as for polytopes;   these examples are  
highly simplified models of what is done for nuclear fusion in
tokamacs. We also present  some
open problems.

\end{abstract}

\tableofcontents

\section{Introduction}
\subsection{The problem}
Let us consider a particle in a  domain $\Omega$ in $\bR^d $
 ($d \geq 2$) in the presence of a magnetic field $B$.
We will always assume that the topological boundary
$\pa \Omega := \overline{\Omega}\setminus \Omega $ of $\Omega$ is compact.
 At the classical  level, if
the strength  of the field  tends to infinity as $x$
approaches
 the boundary $\partial \Omega$, we expect that
 the charged particle is confined and never
visits the boundary: the Hamiltonian dynamics is complete.
 At the quantum level the fact that the particle
 never feels the boundary amounts to saying that the magnetic
field completely determines the motion, so there is no need for
boundary
 conditions. At the mathematical level,  the problem is to  find
 conditions
  on the behaviour of  $B(x)$  as $x$ tends to $\partial \Omega$
 which ensure that the magnetic
operator $H_A$ is essentially self-adjoint
(see Section \ref{sec:ess-self})  on $C_0^{\infty}(\Omega)$
(the space of compactly supported smooth functions).
 These conditions will  not depend on the gauge $A$, but only on the
 field $B$.
One could have called such pairs $(\Omega ,A)$  ``magnetic bottles'',
but this denomination is already introduced in the important
paper \cite{Avr} for Schr\"odinger operators with magnetic fields in
the whole of $\bR^d$
having  compact resolvents.
This question may be  of technological interest in the construction
of tokamacs for the nuclear fusion \cite{toka}. The ionised plasma 
which is heated
is confined thanks to magnetic fields. 

\subsection{ Previous works}
The same problem, concerning scalar (electric) potentials, has been intensively
studied. 
In the many-dimensional case the basic result appears in a paper by
B. Simon \cite{Si} which generalises results of H. Kalf, J. Walter and
U.-V. Schminke (see \cite{KSWW} for a general review). 
Concerning the magnetic potential, the first general
result is by  Ikebe and Kato:
in \cite{IK}, they prove 
 self-adjointness in the case of $\Omega =\bR^d$ for any
regular enough magnetic potential. This result was then improved in 
\cite{Si1,Shu}.
Concerning domains with boundary, we have not seen results in the
purely magnetic case. 
A regularity condition on the direction of the magnetic field
was introduced  in the important paper \cite{Avr}
(Corollary 2.10 p. 853) in order to
construct ``magnetic bottles'' in $\bR^d$. It was used later in many
papers like \cite{Col1,Du,truc1,truc2, ES, ESo, ESol, BLS}.

 In the recent paper \cite{Nen}, G. Nenciu and I. Nenciu give an optimal
 condition for essential self-adjointness 
 on the electric potential near the boundary of a bounded
smooth domain; they  use Agmon-type results on exponential
 decay of eigenfunctions combined with multidimensional Hardy
 inequalities.

\subsection{ Rough description of our  results}
 As we will see, in the case of a magnetic potential
 the Agmon-type estimates still
 hold,
 whereas the Hardy inequalities cannot be used because there 
is no separation between kinetic and potential
energy.
 Actually the point  
is that we need, to apply the strategy of \cite{Nen}, some lower
 bound on the magnetic quadratic form $h_A$ associated
 with  the magnetic potential $A$.
Our main result is as follows:
under some continuity assumption on the direction of $B(x)$ at 
the boundary, 
for any $\epsilon >0$ and $R>0$, there exists a constant
 $C_{\epsilon ,R} \in \bR $ such that   the quadratic form 
$h_A$ satisfies the quite optimal bound
\begin{equation}\label{whi} \forall u \in C_0^{\infty}(\Omega),~
 h_A(u) \geq (1-\epsilon) \int_{\Omega \cap
  \{ x|~|x|\leq R \}}
 |B|_{\rm sp}\ |u|^2 \ |dx| - C_{\epsilon ,R}  \ \|u\|^2~. 
\end{equation}
Here $|B(x)|_{\rm sp}$ is a suitable norm on the space of 
bi-linear antisymmetric forms on $\bR ^d$, called the
{\it spectral norm}.
This implies that $H_A$ is \ess ~ if there exists  $\eta >0$
so that 
 $|B(x)|_{\rm sp}\geq  
  (1+\eta)D(x)^{-2} $ where   $D$ is
 the distance to the boundary of $\Omega $.

We study then examples in the following cases:
\begin{itemize}
 \item The domain $\Omega $ is a polytope
\item The boundary  $\pa \Omega $ is smooth and the Euler
  characteristic
 $\chi  (\pa \Omega )$ vanishes  (toroidal domain)
    \item The boundary
  $\pa \Omega $ is smooth and the Euler characteristic $\chi
  (\pa \Omega )$ does not vanish (non toroidal domain)
\item The domain $\Omega $ is $ \bR^3 \setminus 0$ and 
the field is a  monopole 
or a  dipole
\item The domain $\Omega $
is the unit disk:  
for any $\epsilon >0$ and $d=2$, we construct
  an  example of a non \ess~ operator $H_A$ 
 with $|B(x)|_{\rm sp}\sim 
(\sqrt{3}/2 - \epsilon) D(x)^{-2}$  showing that our bound
is rather sharp. 
\end{itemize}

\subsection{ Open problems}
The following questions seem to be  quite interesting:
\begin{itemize}
\item  What are the properties of a {\it classical} charged
particle in a confining magnetic box? 
Are almost all trajectories not hitting the boundary?
\item What is the {\it optimal } constant $C$  in the estimates 
$|B(x)|_{\rm sp} \geq C D(x)^{-2}$ of our main result \ref{essd}? From our
main results and the example in the unit disk 
 given in Section \ref{sec:contrex}, we see that
the optimal constant lies  in the interval  $[\sqrt{3}/2, 1]$. 
\end{itemize}

\section{Definitions and background results}

In this section, we will give precise definitions and related
notations.
We will also review some known results with references to the literature.
\subsection{The domain $\Omega$}\label{sec:omega}
In what follows, we will keep the following definitions:
$\Omega $ is an open set  in the Euclidean space  $\bR ^d$  ($d\geq 2$)
 with a compact topological 
 {\it   boundary} $\pa \Omega =\overline{\Omega} \setminus
\Omega $, so that either $\Omega $
 or $\bR^d \setminus \Omega $ is bounded.
\begin{definition} \label{defi:R}
We will denote by $d_R $ the distance defined on $\Omega $ by
the Riemannian metric induced by the Euclidean metric:
\[ d_R (x,y)=\inf_{\gamma \in  \Gamma_{x,y}} {\rm length}(\gamma ) \]
where $ \Gamma_{x,y}$ is the set of smooth curves $\gamma :[0,1]
\ra \Omega $ with $\gamma (0)=x,~\gamma (1)=y$.

We will denote by $\widehat{\Omega }$ the metric completion
of $(\Omega, d_R)$ and by $\Omega _\infty=\widehat{\Omega }
\setminus \Omega $ the metric boundary of $\Omega $.

We say that $\Omega $ is {\rm regular }
if $\Omega _\infty$ is compact.

\end{definition}

 If $\Omega  $ is regular, $  \pa \Omega $ is compact.
In fact the identity map of $\Omega $ extends to a continuous
map $\pi $  from $\widehat{\Omega } $ onto $\overline{\Omega }$ and 
$\pi(\Omega _\infty)= \pa \Omega $.
 $(\widehat{\Omega },\pi)$ is a ``desingularisation'' of $\overline{\Omega }$.
If  $X= \pa \Omega $ is 
a  compact $C^1$ sub-manifold or    a
 compact simplicial complex embedded in a piecewise $C^1$ way,
$\Omega $ is regular.

If $X=\cup _{n \in \bN}[0,1]e_n $
with $e_n$ a sequence of unit vectors in $\bR^2$ converging
to $e_0$, then $\bR^2 \setminus X$ is not regular, even
if $\pa \Omega =X$ is compact.

\begin{figure}[hbtp]
    \leavevmode \center 
\input{etoile.pstex_t}
\caption{An example where $\pa X$ is compact while
$X_\infty $ is not compact }
   \label{fig:etoile}
\end{figure}
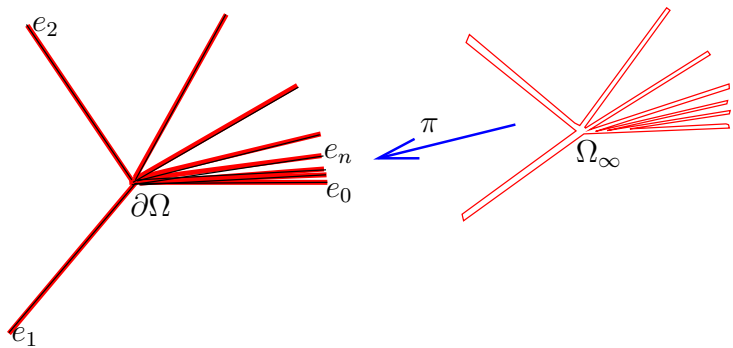
We will use the following regularity property:
\begin{definition}\label{defi:regular} Let us assume that
$\Omega $ is regular. 
 A continuous function $f:\Omega \ra \bC $ is 
{\rm regular at the boundary} if
it extends by continuity to $\widehat{\Omega }$.
\end{definition}

The Lebesgue measure will be $|dx|$ and
we will  denote by 
$\langle u, v\rangle :=\int _\Omega u\bar{v} |dx|$
the $L^2$ scalar product
and  by $\| u\|$  the $L^2$ norm of $u$.
We will denote by $C_0^\infty (\Omega) $ the space of complex-valued
smooth functions with compact support in $\Omega $.

\subsection{The distance to the boundary}

\subsubsection{The distance function}

\begin{definition}
Let us denote by $\hat{d}_R$ the
extension of  $d_R$ by continuity to $\widehat{\Omega}$.
 For
 $x \in \Omega$,
let   $D(x)$ be  the    distance to the boundary
$\Omega _\infty $, given by 
$D(x) = \min _{y\in  \Omega_\infty  }\hat{d}_R(x,y) $.
\end{definition}
\begin{lemma}\label{lemm:lip}
The function $D$ is 1-Lipschitz and  almost everywhere
differentiable in $\Omega $. At any point $x$ of differentiability
of $D$, we have 
$|dD (x)| \leq 1$.
\end{lemma}
The inequality $|D(x)-D(x')|\leq d_R(x,x')$ follows from
the triangle  inequality for $\hat{d}_R$. 
The almost everywhere differentiability of Lipschitz functions
is the celebrated Theorem of Hans  Rademacher \cite{rad1}; see 
also \cite{Morrey} p. 65 and 
 \cite{Hei}.

\subsubsection{ Adapted charts for smooth boundaries}\label{charts}
Assuming that the boundary is smooth, 
we can find, for  each point $x_0\in \pa \Omega $, 
a  diffeomorphism from an open neighbourhood $U$ of $x_0 $
in $\bR ^d$ onto an  open neighbourhood $V$ of $0 $
in $\bR ^d_{x_1,x'}$  satisfying:
\begin{itemize}
\item $x_1 ( F(x) )= D(x)$
\item The differential $F'(x_0)$ of $F$ is an isometry
\item $F(U\cap \Omega )= V\cap \{ x_1 >0 \}$.
\end{itemize}
We will call such a chart an adapted chart at the point $x_0$.
Such a chart is an $\epsilon -$quasi-isometry (see the definition in 
Section \ref{appC})
with $\epsilon$ as  small 
 as one wants  by choosing $U$ small enough.

\subsection{Antisymmetric  forms} \label{sec:anti}

Let us denote by $\wedge ^k \bR ^d$
the space of real-valued k-linear antisymmetric forms on the
 Euclidean space    $\bR^d$.
The space $\wedge ^1 \bR ^d$ is the dual of $\bR^d$, and it is
equipped with
  the natural Euclidean norm: $|\sum_{j=1}^d  a_j dx_j |^2=\sum_{j=1}^d  a_j^2 $.
The space $\wedge ^2 \bR ^d$ is equipped with the spectral
norm: if $B\in \wedge ^2 \bR ^d$,
 there exists an orthonormal basis
of $\bR^d$ so that
$B=b_{12} dx_1 \wedge dx_2 +b_{34} dx_3 \wedge dx_4 +\cdots +
b_{2\bar{d}-1,2\bar{d}}  $  
         with $\bar{d}=[d/2]$
and  $b_{12} \geq b_{34} \geq \cdots \geq  0$; the sequence
$b_{12},  b_{34}, \cdots $ is unique: the  eigenvalues
of the antisymmetric endomorphism  $\tilde{B}$ of
$\bR^d$  associated with  $B(x)$
are  $\pm i  b_{12},\pm i b_{34}, \cdots, \pm ib_{2\bar{d}-1,2\bar{d}}
$
and $0$ if $d$ is odd.
\begin{definition}\label{defi:normB}
We define  the {\rm spectral norm}
 of $B$ by  $|B|_{\rm sp} :=\sum_{j=1}^{\bar{d}}  b_{2j-1,2j} $.
 \end{definition}
  $|B|_{\rm sp}$ is  one half of the
 trace norm of  $\tilde{B}$, hence $|B|_{\rm sp}$  is a norm
on $\wedge ^2 \bR ^d$.
If $d=2$,  $|B|_{\rm sp}=|B|$;
  if $d=3$, $|B|_{\rm sp}$ is the
Euclidean norm of the vector field $\vec{B}$ associated with  $B$, 
defined by $\iota (\vec{B}) dx\wedge dy \wedge dz =B$
where $ \iota (\vec{B})\omega  $ is the {\it inner product} of the 
vector field $ \vec{B}$ with the differential form $\omega $.
\begin{remark} \label{rem:inf-spectre}
 $|B|_{\rm sp}$  is  the infimum
of  the spectrum of the Schr\"odinger operator
with constant magnetic field $B$ in $\bR ^d$.
\end{remark}

\subsection{Magnetic fields}

Let us give the basic definitions and notations concerning
magnetic fields in a domain $\Omega $.
The {\it magnetic potential} is 
a  smooth real 1-form $A$  on
 $\Omega \subset \bR^d$, given by  $A=\sum_{j=1}^d  a_j dx_j$,
 and  the associated 
 {\it magnetic field}  is  the 2-form $B=dA$; more explicitly,
we have $B(x)= \sum_{ 1\leq j<k \leq d} b_{jk}(x) dx_j\wedge dx_k$ 
 with $ b_{jk}(x) = \partial_j a_k(x)-\partial_k a_j(x)\ .$


Let us define now the Schr\"odinger operator with magnetic field
$B=dA$: 
\begin{definition}
The {\rm magnetic connection} $\nabla =(\nabla _j)$ is the differential 
operator defined
by
\[ \nabla _j=
\frac{\pa }{\pa x_j } -i a_j  ~.\]
The {\rm magnetic Schr\"odinger} operator $H_A$ is defined
by
 \[H_A =-\sum_{j=1}^d \nabla _j^2 ~.\]
The {\rm magnetic Dirichlet integral } $h_A=\langle H_A .| . \rangle $
is given, for $u\in C_0^\infty (\Omega )$,
by
\[ h_A (u)= \int _\Omega \sum_{j=1}^d |\nabla _j u|^2 |dx| ~.\]
\end{definition}

Let us note the    commutator formula 
$ [\nabla_j,\nabla _k]= -i b_{jk}$ 
which will be very important. 
From the previous definitions and the fact that the formal adjoint
of $\nabla _j$ is $-\nabla _j$, it is
clear that the operator $H_A $ is  symmetric
on $C_0^\infty (\Omega)$. 
\begin{definition}
 We  will  say that $B=dA$ is a {\rm confining} field in $\Omega $
if $H_A $ is \ess ~ (see Section \ref{sec:ess-self}).
\end{definition}

\subsection{The Riemannian context}
\subsubsection{``Regular'' Riemannian manifolds }
The context of an Euclidean domain is not the most natural one for our
problem. In particular, the ``regularity assumption'' of Definition
\ref{defi:R} can easily be extended to 
the Riemannian context.
Let  $(\Omega ,g) $ be  a smooth Riemannian manifold.
We are interested in cases where $(\Omega ,g) $ is not complete.
Let us recall that $g$ induces on $\Omega $ a distance $d_g$
defined by 
$ d_g(x,y)= \inf _{\gamma  \in \Gamma _{x,y}} {\rm length}(\gamma ) $
where $ \Gamma _{x,y}$ is the set of smooth paths
$\gamma :[0,1 ] \ra \Omega $ so that $\gamma (0)=x,~\gamma (1)=y $.
 We will denote by $\widehat{\Omega }$ the {\it metric 
completion} of $\Omega $ and by $\Omega_\infty  =\widehat{\Omega
}\setminus
\Omega $ the metric boundary.
In the case where $\Omega \subset \bR^d$ is equipped with 
the Euclidean Riemannian metric,  $\Omega_\infty $ is in general
not equal to the boundary $\pa \Omega $.

Definition \ref{defi:R} is now replaced by:
\begin{definition} \label{defi:Rg}
The Riemannian manifold $(\Omega, g)$ is {\rm regular }
if
\begin{enumerate}
\item $\Omega _\infty $ is compact
\item For any $\epsilon >0$, 
every  $x_0 \in \Omega _\infty $ has a neighbourhood $U$
so that so that  $U \cap \Omega $
is $\epsilon -$quasi-isometric (see Definition \ref{defi:quasi}) to an
open set of $\bR^d$ with an Euclidean metric.
\end{enumerate}
A function $f:\Omega \ra \bC $ is {\rm regular} at the boundary 
if it extends by continuity to $\widehat{\Omega }$.
\end{definition}


\subsubsection{Magnetic fields on Riemannian
manifolds}

The magnetic potential is a smooth real valued 1-form $A$ on $\Omega $,
the magnetic field is the 2-form $B=dA$. The norm
$| B(x) |_{\rm sp}$ is calculated with respect to the
Euclidean metric $g_{x_0}$. 
The magnetic potential defines a connection $\nabla $  on the trivial line
bundle $\Omega \times \bC \ra \Omega $
by $\nabla _X f=d f(X)  -iA f $. 
The magnetic Dirichlet integral is
$h_A (f)=\int _\Omega \| \nabla f \| ^2_g |dx|_g $
where the norm of the 1-form $\nabla f(x) $ is calculated 
with the dual Riemannian norm:
 $\| \nabla f \| ^2_g=\sum _{ij}g^{ij}\nabla _{\pa _i}f\nabla _{\pa _j
 }f$
and $|dx|_g=\theta |dx_1 \cdots dx_d| $ is the Riemannian volume.
The magnetic Schr\"odinger operator is then 
defined by:
\[ H_A f= -\theta^{-1}\sum _{ij} \nabla _{\pa _i}\left(
\theta g^{ij} \nabla _{\pa _j} f \right)~.\] 

\subsection{Essential self-adjointness} \label{sec:ess-self}
In this section, we will review what is an essentially self-adjoint
operator and give some easy propositions which we were not able 
to point in the literature. 
\subsubsection{Essentially self-adjoint operators}
Let us recall
the following 
\begin{definition}
 A differential operator $P:C_0^\infty (\Omega )
\ra C_0^\infty (\Omega )$ is {\rm  essentially self-adjoint}
in $L^2(\Omega ,|dx|)$  if $P$ is
formally symmetric (for any $u,v \in C_0^\infty (\Omega )$,
$\langle Pu | v \rangle =\langle u |P v \rangle  $)
and the closure of $P$ is self-adjoint.
\end{definition}

A basic criterion  for essential self-adjointness is   the following
(see criterion (4) of Theorem X.1 and Corollaries  in  \cite{RS} ):
\begin{proposition} \label{prop:ess-sa} 
 Let $P$ be as before and formally symmetric.
Let us assume either that 

(1) 
 there exists $E \in \bR $  so that any solution 
$v\in L^2 (\Omega )$ of  $ (P -E) v= 0 $ (in the weak sense 
of Schwartz distributions)
vanishes,

or that

(2) there exists  $\lambda_\pm \in \bC $ with $\pm  \Im  (\lambda_\pm ) >0$
so that any solution 
$v\in L^2 (\Omega )$ of  $ (P -\lambda_\pm ) v= 0 $ (in the weak sense 
of Schwartz distributions)
vanishes.

Then $P$ is essentially self-adjoint.

\end{proposition}
\subsubsection{Essential self-adjointness  depends only on the
boundary behaviour}

\begin{proposition}
Let $X$ be a smooth manifold with a smooth density $|dx|$. Let
$L_j,~j=1, 2$ be two  formally symmetric elliptic differential  operators 
of degree 
$m$ on $L^2(X,|dx|)$ and 
 let us  assume that $L_1 $ is \ess~ and $L_2-L_1 =M $ is
compactly
 supported.
Then $L_2 $ is \ess.
\end{proposition}
{\noindent {\it Proof.--}}
It is enough to show that $L_2 -ci $ is invertible
for $c $ real and large enough.
We have
$L_2 -ci= \left( {\rm Id}+M(L_1-ci)^{-1}   \right)
(L_1-ci) $. Moreover the domain of $L_1$ contains  $H^m _0$
 (the space of compactly supported $H^m$ functions).
So that $\|M(L_1-ci)^{-1} \|=O(c^{-1})$. 
\hfill $\square$

\subsubsection{Essential self-adjointness is independent of
the choice of a gauge}

\begin{proposition}
 Let $X$ be a smooth manifold with a smooth density $|dx|$.
Let us consider a Schr\"odinger operator $H_{A_1}$ and 
$A_2=A_1 + dF $ with $F\in C^\infty (X,\bR)$.
 Then,  if $H_{A_1}$ is \ess,  $H_{A_2}$ is
also  \ess.
\end{proposition}
{\noindent {\it Proof.--}} We have formally (as differential operators)
\[ H_{A_2}=e^{iF} H_{A_1}e^{-iF}~.\]
Hence, $ H_{A_2}-ci =e^{iF} \left( H_{A_1}-ci \right) e^{-iF}$.
The domain $D_2$ of the closure of $H_{A_2}$ (defined on
$C_0^\infty (X)$)
is $e^{iF}$ times the domain $D_1$ of the closure of $H_{A_1}$.
The result follows from the fact that  $e^{\pm iF}$
is invertible in $L^2$ and an isomorphism of the domains.
 \hfill $\square$
\section{Main results}

Let us take $H_A$ 
 with domain ${\cal D} (H_A) =C_0^{\infty}(\Omega)$.
 As explained in the introduction, we are looking for
 growth assumptions on  $|B|_{\rm sp}$
close to $\partial \Omega$ ensuring essential self-adjointness of $H_A$.
We formulate now our main results:
\begin{theorem}\label{ess2}
Let us take $d=2$.
Assume that $\pa \Omega $ is compact with
a finite number of connected components  and that 
$B(x)$ satisfies near $\pa \Omega $
\begin{equation}\label{bune}
 |B(x)|_{\rm sp}  \geq (D(x))^{-2}~, 
\end{equation}
then the Schr\"{o}dinger operator $ H_A$ is essentially self-adjoint.
This  still holds true for  any gauge $A'$ such that $dA'=dA=B$.

\end{theorem}

\begin{theorem}\label{essd}
 Let us take $d>2$.
Assume that $\Omega $ is regular and 
  that 
 there exists $\eta >0$ 
such that
$B(x)$ satisfies near $\pa \Omega $
\begin{equation}\label{bun}
|B(x)|_{\rm sp} \geq  \left( 1   + \eta  \right)
  (D(x))^{-2}~,
\end{equation}
and that   the functions 
\begin{equation}\label{bunn}n_{jk}(x) = \frac{b_{jk}(x)}{|B(x)|_{\rm sp}}   
 \end{equation}
are regular at the
boundary  $ \Omega_\infty  $ 
(for any $1\leq j<k \leq d$) (see Definition \ref{defi:regular}).
Then the Schr\"{o}dinger operator $ H_A$ is essentially self-adjoint.
This  still holds true for  any gauge $A'$ such that $dA'=dA=B$. 
\end{theorem}
\begin{remark}
If $\Omega $ is defined (locally or globally) by
$\Omega :=\{ x\in \bR^d~|~f(x)>0 \}$ with $f:\bR  ^d \rightarrow \bR $
smooth, 
$df (y)\ne 0 $ for $y\in\pa \Omega $, then
$f(x)\sim |df (x)|D(x)$ for $x$ close to $ \pa \Omega $.
And we can replace in the estimates (\ref{bun}) 
$D(x)$ by $f(x)/ |df (x)|$. 
\end{remark}

Theorem \ref{essd} can be extended to Riemannian manifolds as
follows:
\begin{theorem} \label{theo:ess-riem}
Let $(\Omega ,g)$ be a regular Riemannian manifold
with a magnetic field $B=dA $.
Let us assume that 
$\| B \| _{\rm sp} \geq (1+\epsilon ) D^{-2 } $
near $\Omega _\infty $
and that, for each $x_0\in\Omega _\infty$,
the direction  $n(x) $  of $B$, calculated with the metric $g_{x_0}$
(i.e. using the trivialisation of the tangent bundle associated with  
$g_{x_0}$),
has a limit as $x\ra x_0$,
then  $H_A $ is \ess ~ on $C_0^\infty (\Omega )$. 
\end{theorem}

 The exponent $2$ of the leading term in Equations (\ref{bune})
 and (\ref{bun}) is optimal, as shown  in the following
\begin{proposition}\label{contrex}
For any $0<\alpha<\sqrt{3}/2$, there exists a magnetic
 field $B$ such that $H_A$ (with $dA=B$) is not essentially self-adjoint
 and such that the growth of 
$|B|_{\rm sp}$ near the boundary $\partial \Omega$ satisfies
$$|B(x)|_{\rm sp} \geq \  \ \frac{\alpha}{(D(x))^2} \quad  .$$
\end{proposition}
We prove this proposition in Section \ref{sec:contrex} in the case $d=2$,
 but the proof
 can be easily generalised to larger dimensions.

As a consequence of this proposition, together with Theorem 
\ref{ess2} (respectively  \ref{essd} ), we get   that the optimal
 constant in front of the leading term 
$(D(x))^{-2}$ is in  $[\sqrt{3}/2,  1]$.

Hence we see that the situation for confining magnetic fields is not
 the same as for confining potentials (for which the optimal constant
is $3/4$, hence is smaller than $\sqrt{3}/2$).

Indeed this is due to the difference between the Hardy inequalities in
 the two situations: the term $1/(4 D^{2})$ does not appear in
 the magnetic case, as it does in the
case of a scalar potential, where it plays the role of an "additional barrier".

\section{Proof of the main results}

In this Section, we prove Theorems \ref{ess2}, \ref{essd}
and \ref{theo:ess-riem} using the method of \cite{Nen} which we first
review.

\subsection{Agmon  estimates}

 The following statement is  proved, using  Agmon estimates \cite{Agm},
in  \cite{Nen}:
\begin{theorem}\label{Hur} Assume that $\pa \Omega $ is compact, 
and that there exists  $c\in \bR $ such that,
 for all $u \in C_0^{\infty}(\Omega)$,
$h_A(u)  - \int_{\Omega} D(x)^{-2}
|u(x)|^2
 |dx| \geq c \|u \|^2$. Then 
 $H_A$ is \ess. 
\end{theorem}

Reading the proof in \cite{Nen}, one sees that the only property
of $\Omega $ which is used is that the function $D(x)$ is
smooth near the boundary and satisfies $|dD(x)| \leq 1$.
One can extend the proof to the case where $\pa \Omega $ is not
a smooth manifold by  using the properties  of the function $D$ described
in Lemma \ref{lemm:lip}. The fact that $\Omega $ is bounded does not
play an 
important role,
only the compactness of $\pa \Omega $  is important. 
The essential self-adjointness of $H_A $ results from   the  Proposition
\ref{prop:ess-sa} and the following

\begin{theorem}\label{Hor}
Let $v\in L^2(\Omega ) $ be a weak solution of $(H_A-E)v=0$.
Let us assume that there exists
  a constant $c>0$ such that, for all $u \in C_0^{\infty}(\Omega)$, 
\begin{equation} \label{equ:assump}
\langle u | (H_A-E) u \rangle -
 \int_{ \{x\in \Omega ~|~ D(x)\leq 1 \}}
 \frac{ |u(x)|^2}{D(x)^2} |dx|_g \geq c \|u\|^2~. \end{equation}
Then $v\equiv 0$.
\end{theorem}
For the reader's convenience, we give here the proof
of Theorem \ref{Hor} following  the strategy of \cite{Nen} in a slightly simplified way.

{\noindent {\it Proof.--}}
The proof  is based on the following simple identity (\cite{Nen})
\begin{lemma}\label{lemma:ute}
Let $v$
 be a weak solution of $(H_A-E)v=0$, and let $f$
be a real-valued Lipschitz function with compact support.
 Then
 \begin{equation}\label{ute}
\langle   fv | (H_A-E) (fv) \rangle   =
  \langle   v~ |~ |d f(x)|^2  v \rangle ~.   
\end{equation}
\end{lemma}

Let us give two numbers $\rho $ and $R$ satisfying respectively 
$0< \rho < \frac{1}{2}$
and $ 1 < R < +\infty$.
We will apply identity (\ref{ute})
with $f=F(D) $ where $F(u)$ the piecewise smooth function defined by
\[ F(u)= \left\{ 
\begin{array}{l}
0  {\rm ~ for~}  u\leq \rho 
{\rm ~and~ for~ }  u \geq R+1  \\
 2(u-\rho )   {\rm ~ for~} \rho \leq u \leq 2 \rho  \\
u   {\rm ~ for~} 2 \rho \leq u \leq 1  \\
1   {\rm ~ for~} 1 \leq u \leq R  \\
 R+1 -u   {\rm ~ for~} R \leq u \leq R+1   
\end{array}
\right.
\]
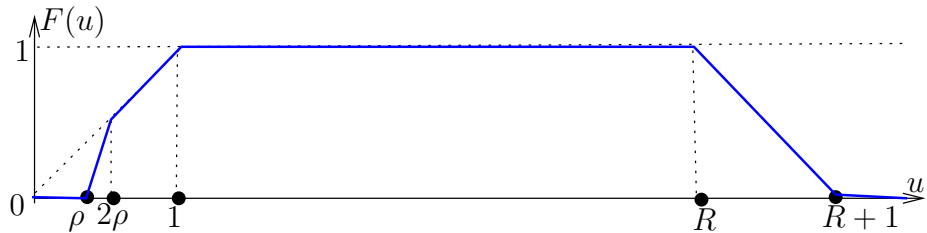
\begin{figure}[hbtp]
    \leavevmode \center 
\input{functionF.pstex_t}
\caption{The function $F$}
   \label{fig:functionF}
\end{figure}

We have $|d f|^2 = F'(D)^2$ almost everywhere.
From the inequality  (\ref{equ:assump}) applied to
$fv$, we get:
\begin{equation}\label{equ:I}
\langle (H_A-E)(fv)~|~fv \rangle \geq
\int_{ 2\rho \leq  D(x) \leq 1 } |v|^2 |dx|_g +c \| fv\| ^2 ~.
\end{equation}
On the other hand, using the explicit values of $df$
and Equation (\ref{ute}),
we get:
\begin{equation}\label{equ:II} 
\begin{array}{rl}
\langle (H_A-E)(fv)~|~fv \rangle \leq& 4\int_{ \rho \leq  D(x) \leq
  2\rho } |v|^2|dx|_g + \cdots \\
\cdots  \int_{ 2\rho \leq  D(x) \leq 1 } |v|^2 |dx|_g
+ & \int_{ R \leq  D(x) \leq R+1 } |v|^2 |dx|_g ~.
\end{array}
\end{equation}
Putting together the inequalities (\ref{equ:I}) 
and (\ref{equ:II}), we get
\begin{equation}\label{equ:III} c \| fv\| ^2
\leq 4\int_{ \rho \leq  D(x) \leq
  2\rho } |v|^2|dx|_g +  \int_{ R \leq  D(x) \leq R+1 } |v|^2 |dx|_g
~. 
\end{equation} 
Taking $\rho \ra 0 $ and $R\ra +\infty $ in the inequalities
(\ref{equ:III}),
 we get that
the $L^2$ norm of $ v$ vanishes.

\hfill $\square$

\subsection{Quasi-isometries}\label{appC}
In section 5 we  give examples which have smooth boundaries
 (excepting the convex polyhedra (section \ref{sec:poly})).
In order to build new examples, like  non convex
polyhedra, one can 
 use quasi-isometries.

 \begin{definition}\label{defi:quasi} Given $0<c \leq C $, 
a $(c,C)$-quasi-isometry of $\Omega _1 $ onto
$\Omega _2 $ is an homeomorphism
of $F:\overline{\Omega _1}$ onto  $\overline{\Omega _2}$
whose restriction to $\Omega _1$ is a smooth diffeomorphism
onto $\Omega _2$ and such that 
\[ \forall x,y \in \overline{\Omega _1}, ~cd_R(x,y)\leq d_R(F(x),F(y))
\leq Cd_R(x,y) ~.\]

An $\epsilon - $quasi-isometry is an $(1-\epsilon, 1 +\epsilon)$
quasi-isometry.
\end{definition}
\begin{lemma}\label{lemm:bounds}
We have the bounds

$ \| F' \| \leq C,~ \| (F^{-1})' \| \leq c^{-1} $, 
$ |{\rm det}(F')|\leq C^d $ ,
$ cD_1 (x)  \leq D_2 (F(x)) \leq CD_1 (x)$,
where, for $i=1,2$,    $D_i(x)$ denotes, for
any $x \in \Omega_i$, the   distances to the boundary
$ (\Omega_i)_\infty $.

\end{lemma}
We will start with a magnetic potential $A_2$ in $\Omega _2$
and define $A_1=F^\star (A_2)$.
We want to compare the magnetic quadratic forms
$h_{A_2} (u) $ and $h_{A_1}( u\circ F) $ as well as the $L^2$
norms. 
 We get:
\begin{theorem}\label{theo:quasi}Assuming that,
for any $u\in C_0^\infty (\Omega _2), $
\[ h_{A_2}(u)\geq K\int_{\Omega _2}\frac{|u|^2}{D_2^2}|dx_2| -L \|u
\| ^2 ~,\]
we have, for any   $v\in C_0^\infty (\Omega _1)$, 
\[ h_{A_1}(v)\geq K\left( \frac{c}{C} \right)^{d+2}
\int_{\Omega _1}\frac{|v|^2}{D_1^2}|dx_1| -{L}{c^2} \| v  
\| ^2 ~.\]
\end{theorem}
In other words, we can check that $H_{A_1}$ is \ess ~ from
an estimate for $h_{A_2}$ using Theorem \ref{Hur}. 

{\noindent {\it Proof.--}}
Let us start making the change of variables
$x_2=F(x_1)$ in the integral $h_{A_2}(u)$. 
Putting $v=u\circ F $, we get
$h_{A_2}(u)=\int _{\Omega _1} \| \nabla _{A_1}v(x_1)\| ^2_g |{\rm det 
     }(F'(x_1))||dx_1 | $ where $g$ is the inverse of the
pull-back of the Euclidean metric by $F$.  
Using Lemma \ref{lemm:bounds}, we get the estimate. 
\hfill $\square$

\subsection{Lower bounds
for the magnetic Dirichlet integrals}

\subsubsection{Basic magnetic estimates}

\begin{lemma} \label{lemm:general}
For any $u\in C_0 ^\infty (\Omega )$, we have
\[ h_A(u) \geq | \langle b_{12} u|u \rangle |
+  | \langle b_{34} u|u \rangle | + \cdots +
 | \langle b_{2\bar{d}-1,2\bar{d}} u|u \rangle |  ~.\] 
\end{lemma}
{\noindent {\it Proof.--}}
We have
\[| \langle b_{12} u|u \rangle |  =|\langle [\nabla _1 ,\nabla_2 ]u|u
\rangle| \leq 2 | \langle \nabla _1 u |\nabla _2 u \rangle|\leq
\int_\Omega (|\nabla _1 u| ^2 +|\nabla _2 u| ^2)|dx|   ~.  \]
We take the sum of similar inequalities replacing
the indices $(1,2)$ by $(3,4),\cdots,(2\bar{d}-1,2\bar{d})  $. 
\hfill $\square$
\begin{lemma} \label{lemm:main-estim} Let $\Omega $ be a regular open
set in $\bR ^d$. 
Let $x_0 \in  {\Omega }_\infty  $ and assume that $B(x)$ does not vanish
 near the point $x_0$ and that the direction of $B$ is regular 
near $x_0$.
Let $A$ be a local potential for $B$ near $x_0$, then, for any
$\epsilon >0$, there exists a neighbourhood $U$ of $x_0 $
in $ \bR ^d$ so that, for any $\phi \in C_0^\infty (U\cap \Omega )$,
\begin{equation}\label{estimate}
 h_A (\phi )\geq (1-\epsilon)\int _U |B(x)|_{\rm sp} |\phi (x)|^2 |dx| ~,
\end{equation}
where  $|B(x)|_{\rm sp} $ is defined in Definition \ref{defi:normB}.
\end{lemma}
{\noindent {\it Proof.--}}
Let us choose  
 $U$ so that, for all
$x\in U\cap \Omega $,  $|n (x) - n (x_0)|_{\rm Eucl} \leq
  \epsilon\sqrt{\frac{2}{d(d-1)}} $, where 
$|\sum_{i<j} a_{ij}dx_i \wedge dx_j|_{\rm Eucl}^2=\sum_{i<j} a_{ij}^2$,
  by applying Definition \ref{defi:regular} to $n(x)$ 
   at the point $x_0$.
We choose orthonormal coordinates in $\bR ^d $ so
that $n (x_0)=n_{12}  dx_1 \wedge dx_2 +
 n_{34} dx_3 \wedge dx_4 +\cdots $
with $n_{2k-1,2k}  \geq 0 $ and $\sum_k  n_{2k-1,2k} =1$. 
 From Lemma \ref{lemm:general}, we have, for $\phi \in C_0^\infty
(\Omega \cap U)$, 
 \[h_A (\phi )\geq   \int_{U} |B(x)|_{\rm sp}
({n_{12}  (x) +n_{34}(x) + \cdots }) |\phi (x)
 |^2|dx|~\]
and $n_{12}  (x) +n_{34}(x) + \cdots \geq 1-\epsilon$,
because the Euclidean norm of $n(x)$ is independent of the
orthonormal basis. \hfill $\square$
\begin{remark}The estimate (\ref{estimate}) is optimal
in view of  Remark \ref{rem:inf-spectre}.
\end{remark}

\subsubsection{The 2-dimensional case}\label{ss:hardy2d}
\begin{theorem} \label{Har2}
Let us assume that $\pa \Omega \subset B(O,R)$ and that $\pa  \Omega$
has a finite number of connected components.
 If $d=2$ and  if $B$ does not vanish near  $\pa \Omega$, then 
there exists $c_R\in \bR$ so that,  $\forall u \in C_0^{\infty}(\Omega)$,
 \begin{equation}\label{H2} 
h_A(u) \geq  \int_{\Omega\cap B(O,R)}|B| |u|^2  |dx|
-c_R  \| u\| ^2~.
\end{equation}
\end{theorem}
{\noindent {\it Proof.--}}
As $B$ does not vanish  near $\pa \Omega $, the sign of $B$ is constant 
near each connected component of $\pa \Omega $.
Let us write 
$\overline{\Omega} \subset  \cup_{l=1}^3 \Omega _l $ with $\Omega _l$ 
open sets
 such that  $\Omega _1 \cap \pa \Omega =\emptyset $, $B>0$  
 on $\Omega _2$ and $B<0$ on $\Omega _3$.
 We can assume that $\Omega _2$ and $\Omega _3$ are bounded. 
Take a partition of unity $\phi_j,~j=1,2,3, $
so that, for $j=2,3$,   $\phi_j \in C_0^\infty (\Omega _j)$,
and $\sum \phi_j  ^2 \equiv 1$.

Now we use the IMS formula (see  \cite{sigal})
\begin{equation} \label{equ:IMS}
h_A(u)=\ \sum_{l=0 }^{2} h_A(\phi_l u)\ -
\int_{\Omega}\left( \sum_{l=0 }^{2} |d \phi_l|^2\right) |u|^2\ |dx|~.
\end{equation}
 with  the lower bound of Lemma \ref{lemm:general}
    in  $\Omega _l \cap \Omega $ for $l=2,3$ and the lower bound
$0$ for $\Omega _1$.
\hfill $\square$

\subsubsection{The  case $d>2$}\label{sec:ims}

\begin{theorem}\label{Har}
Let us assume that $\pa \Omega \subset B(O,R)$.
Assume that $B=dA$ does not vanish
near $\pa \Omega $  and that the functions
$n_{jk}(x)$ are regular at the boundary  $\pa \Omega $,  
then, for any $\epsilon >0$, there exists $C_{\epsilon ,R} >0$
so that, $\forall u \in C_0^{\infty}(\Omega)$,
\begin{equation} \label{equ:hardy}
 h_A(u) \geq (1-\epsilon) \int_{\Omega \cap B(O,R)} |B|_{\rm sp} 
 |u|^2  |dx| - C_{\epsilon , R}  \int_{\Omega} |u|^2  |dx| ~.
\end{equation}
\end{theorem}

{\noindent {\it Proof.--}}
We first choose a finite covering of $ \Omega_\infty  $ by open sets
$U_l,~l=1,\cdots N$ of
$\bR^d$  which satisfies the estimates of Lemma
\ref{lemm:main-estim}.
We choose then a partition of unity 
$\phi_l , l=0,\cdots , N$
with 
\begin{itemize}
\item  For $l\geq 1 $, $\phi_l \in C_0^\infty ( U_l)
  $
\item $\phi_0 $ is $C_0^\infty(\Omega) $
\item $ \sum_l \phi_l ^2 \equiv 1 $ in $\Omega $
\item $\sup \sum _l |d\phi_l|^2 =M$.
\end   {itemize}

 Using the estimates given
in Lemma \ref{lemm:main-estim} for $l\geq 1$   and the
fact that   $\sum_l  |d\phi_l|^2 $ is bounded by $M$, we get,
using IMS identity (\ref{equ:IMS}), 
the inequality (\ref{equ:hardy}).

\hfill $\square$

\subsection{End of the proof of the main theorems}

Using Theorem \ref{Hur}, it is enough to show  
that there exists $c \in \bR$ such that,
for all $u \in C_0^{\infty}(\Omega)$, 
$$h_A(u) \geq  \int_{\Omega\cap B(O,R)} |D(x)|^{-2} |u(x)|^2 |dx| - c \|u\|^2, $$
under  the assumptions of 
Theorems \ref{ess2} and \ref{essd}.
This is a consequence  of Theorem \ref{Har2}  for $d=2$ and Theorem
\ref{Har} for $d>2$.

The proof of Theorem \ref{theo:ess-riem} 
 is an adaptation of the case of an Euclidean domain.
The partition of unity is  constructed using only the distance function 
which has enough regularity.  
We use also the fact that near each point $x_0$ of the boundary
the metric is quasi-isometrically close  to the Euclidean
metric $g_{x_0 }$.
\section{Examples}
\subsection{Polytopes} 
\label{sec:poly}
A {\it polytope} is a convex compact polyhedron.
Let  $\Omega $ be a  polytope given by 
\[ \Omega = \cap _{i=1}^N \{x~|~ L_i(x)  <0 \} ~,\]
where the $L_i $'s are  the  affine real-valued functions
\[ L_i(x)=\sum_{j=1}^d  n_{ij} x_j + a_i~. \]
We will assume that, for 
$ i=1,\cdots,d,~\sum _{j=1}^d n_{ij}^2 =1$ (normalisation)
and $n_{i1 }\ne 0$ (this last condition can always
be satisfied by moving $\Omega $ by a generic isometry).
We have the
\begin{theorem}
The operator $H_A$ in $\Omega $  with 
\[ A= \left( \frac{1}{n_{11}L_1} +\frac{1}{n_{21}L_2} +\cdots  \right) 
  dx_2 ~,\] 
 is \ess.
\end{theorem}
{\noindent {\it Proof.--}}
We have
\[ B= \left( \frac{1}{L_1^2} +\frac{1}{L_2^2} +\cdots  \right) 
dx_1 \wedge   dx_2 + \sum_{j=3}^d  b_j dx_j \wedge dx_2 ~,\] 
and $D=\min_{1 \leq i \leq N} |L_i | $.
So that $B=b_{12} dx_1 \wedge dx_2 +
 \sum_{j=3}^d  b_{j2} dx_j \wedge dx_2 ~  $ with 
$b_{12} \geq D ^{-2}$. 
We then apply directly Lemma \ref{lemm:general}
and Theorem \ref{Hur}. 

\hfill $\square$

\subsection{ Examples in domains  whose
  Euler characteristic of
  the boundary  vanishes (``toroidal domains'').} \label{sec:chine0}

 Let us assume that $\pa \Omega $ is a smooth
compact manifold of co-dimension $1$ and denote 
by $j:\partial\Omega \rightarrow \bR^d$ the injection of $\partial\Omega$ into
$\bR^d$.
A famous theorem of H. Hopf (see \cite{al-ho,gu-po})
asserts that there exists a nowhere vanishing tangent vector field
to $\pa \Omega $
(or 1-form) if and only if the Euler characteristic of 
$\pa \Omega $ vanishes.

\begin{theorem}
Let us assume that the Euler characteristic of $\partial\Omega$
vanishes
(we say that $\Omega $ is toroidal).
Let $A_0$ be a smooth $1-$form on $\overline{\Omega}$
so that the $1-$form on $\pa{\Omega}$ defined by $\omega=j^\star (A_0 )$
 does not vanish, and consider
  a   $1-$form $A$ in $\Omega$ defined, near $\partial\Omega$,    by
$ A= {A_0}/{D^\alpha} $.
We assume that either $\alpha >1$, 
or  $\alpha =1$ with the additional condition that for any $y\in \pa \Omega $,
$| \omega (y)| >1 $.
 Then  $H_A$ is essentially self-adjoint.
\end{theorem}
\begin{remark} The existence of $\omega$ is provided by the topological
assumption on $\pa \Omega $.
This works if $\Omega \subset \bR^3$ is bounded by a 2-torus. It is the
case
 for tokamacs.
\end{remark}

{\noindent {\it Proof.--}}
We will apply Theorem \ref{essd}.
We have  to check:
\begin{itemize}
\item 
 The uniform continuity of the direction 
of the magnetic field or the extension by  continuity to 
$\overline{\Omega}$. It has to be checked locally near the boundary
 $\partial\Omega$. We will use an adapted chart (see section \ref{charts}).

In these local coordinates we write
 $A_0 =a_1 dx _1 + \beta  $ with $\beta = a_2 dx_2 +\cdots + a_d dx_d $
and 
 $\omega = a_2 (0, x') dx _2 + \cdots + a_d(0,x')dx _d  $
so we get
 \[ B=d\left(\frac{A_0}{x_1^\alpha}\right) = \frac{x_1 dA_0
-\alpha dx_1\wedge \beta }{x_1^{\alpha +1}}~. \]
Thus we get that the direction of $B $ 
is equivalent as $x_1 \rightarrow  0^+  $ to that of 
$ dx_1\wedge \omega $
which is non vanishing and continuous on $\overline{\Omega}$.
\item The lower bound (\ref{bun}) $|B|_{\rm sp}\geq (1+\eta )D^{-2}$
near $\pa \Omega $. The norm of $B$ near the boundary
is given, as $x\rightarrow y $ by 
\[ |B(x)|_{\rm sp} \sim \alpha |\omega(y)|/D^{\alpha+1} ~.\]
\end{itemize}
Therefore we conclude that the hypotheses
 of Theorem \ref{essd} are fulfilled.
\hfill $\square$
\begin{remark} The asymptotic
behaviour of $B(x)$ as $x \ra \pa \Omega $
is 
\[ B(x) \sim -\frac{\alpha dx_1 \wedge \omega (y)}{ D^{\alpha +1}}~.\]
It follows that $\omega$ and $\alpha $ depend only of $B$ and 
are invariant by any gauge transform in $ \Omega $. 
\end{remark}

\begin{remark}
If $d=3$, the magnetic field $B$  can be identified with a vector
field $\vec{B}$ in $\Omega $ defined by
\[ \iota \left(\vec{B}\right) dx_1 \wedge dx_2 \wedge dx_3 =B ~\]
as in Section \ref{sec:anti}.
Using the induced Riemannian structure, we can identify any 
1-form $\omega $ on $\pa \Omega $ with a vector field $\vec{\omega }$.
Moreover $\pa \Omega $ is oriented by any 2-form
$\Sigma =\iota (\nu)  dx_1 \wedge dx_2 \wedge dx_3 $
with $\nu $ any outgoing vector field near $\pa \Omega $.
Using the previous identifications,
the asymptotic behaviour of
$\vec{B}$ near $\pa \Omega $ is given by
\[ \vec{B} \sim \alpha r \left( \vec{\omega } \right) / D^{\alpha +1 }
~,\]
where $r$ is the rotation by  $+ \pi /2 $ in the tangent space to
$\pa \Omega $.

It means that $\vec{B}$ is very large near $\pa \Omega $ and 
parallel to  $\pa \Omega $. From the point of view of
classical mechanics,  the
trajectories of the charged particle 
are spiralling around the field lines and do not cross the boundary.
It would be nice to have a precise statement.
\end{remark}

\subsection{Non toroidal domains }
\subsubsection{Statement of results}
We try to follow the same strategy than in Section \ref{sec:chine0},
but now we will allow  the  1-form $\omega $
 on $X=\pa \Omega $ to  have some zeroes. This is forced by the
 topology if the Euler characteristic of $\pa \Omega $ does not
 vanish. 
We need the
\begin{definition}
A 1-form $\omega $ on a compact manifold $X$ is {\rm generic}
if $\omega $ has a finite number of zeroes  and
$d\omega $ does not vanish at the zeroes of $\omega  $.  
\end{definition}
We have the 
\begin{theorem}\label{theo:non-tor}
Let $\Omega \subset \bR^d $ with a smooth compact 
boundary $X=\pa \Omega $. Let $A_0$ be a smooth 1-form
in $\bR^d $ so that $\omega =j^\star_X (A_0)$ is {\rm generic}.
We assume also that, at each zero $m$ of  $\omega $,
\begin{equation} \label{assumption}
|d\omega (m)  |_{\rm sp }>1 ~,\end{equation}
where the norm $ |d\omega (m)  |_{\rm sp }$ is calculated
in the space of anti-symmetric bi-linear forms on the tangent space 
$T_m \pa \Omega $.
Then, if $A$ is a 1-form in $\Omega $ such that   
 near $X$, $A=A_0/D^{2}$,
$B=dA$ is confining in $\Omega $.
\end{theorem}

We see that the field need to be   more singular than in the 
toroidal case. We could have taken this highly singular part     
       only near the zeroes of $\omega $. 
\subsubsection{Local model  }

We will work in an  adapted chart at a zero of $\omega$.
We take  $A= A_0/{x_1}^{2} $ with $j^\star (A_0)=\omega $, we have:
$A_0= a_1  dx_1 + \beta  $
and $\beta   (0)=0 $.

We have 
\[ B= \frac{ d\omega }{x_1^2}  +  
dx_1 \wedge \rho  +0(x_1^{-1}). \]
Applying the basic estimates of Lemma
\ref{lemm:general}  in some orthonormal coordinates
in $\bR ^{d-1}$ so that
$d\omega (0) =b_{23} dx_2\wedge dx_3 + \cdots $,     
 we see, using the assumption  (\ref{assumption}),
 that there exists a neighbourhood $U$ of the
origin and an $\eta >0 $ so that, for any
$u\in C_0^\infty (U)$,
\[ h_A (u) \geq (1+\eta ) \int _U \frac{ |u|^2 }{x_1^2} |dx | ~.\]


\subsubsection{Globalisation}

Near each zero of $\omega $, we take a local chart of $\bR^d$ 
where  $A$ is given by the local
model. Such a chart is an $\epsilon$-quasi-isometry (see \ref{defi:quasi} )   
     with $\epsilon$ as small
as one wants. 
This gives the local estimate near  the 
zeroes of $\omega $. The local estimate    
outside the zeroes of $\omega $ is clear because
we have then $|B|_{\rm sp} \geq  C/D^3 $ with $C>0$: this follows
from the estimates in Section \ref{sec:chine0}  with $\alpha =2$. 
 We finish the
proof of 
Theorem \ref{theo:non-tor} with IMS formula and the local
estimates needed in Theorem \ref{Hur}.

\subsection {An example of a non ~\ess ~Schr\"odinger operator
with large magnetic field near the boundary}
\label{sec:contrex}

Let us consider the 1-form defined on 
 $\Omega =\{(x,y)\in\bR^2|~ x^2+y^2=r^2<1\}$
by    $ A = {\alpha (xdy-ydx)}/({r-1})\ $ where $0< \alpha<
\sqrt{3}/2$. The magnetic  potential $A$ is invariant by rotations.
Then
\begin{theorem} \label{theo:notess}
The operator $H_A$ is not \ess.
\end{theorem}

 The  corresponding magnetic field $B$ writes 
$B(x,y) = \frac{\alpha(r-2)}{(r-1)^2}  dx\wedge dy \ ,$ and,
 near the boundary, 
$|B(x)| \sim  \ {\alpha}/{(D(x))^2}  $.
 We have, in polar coordinates $(r, \theta)$,
$$H_A = -\frac{\partial  ^2}{\partial  r^2}-\frac{1}{r}
 \frac{\partial }{\partial r} -\frac{2i\alpha r}{r-1}\frac{\partial }
{\partial \theta}
+  \frac{\alpha ^2 r^2}{(r-1)^2}\ .$$
Hence the operator $H_A$ splits as a sum
$\sum _{m\in\bZ}H_{A,m}$ where $H_{A,m}$ acts on functions
$e^{im\theta}f(r)$. We will  look at the $m=0$ component: Theorem
\ref{theo:notess} follows from the 
\begin{lemma} If  $0< \alpha<
\sqrt{3}/2$, on the Hilbert space $L^2( ]0,1[, rdr )$, 
the operator 
\[ H= -\frac{d  ^2}{d r^2}-\frac{1}{r}
 \frac{d }{d r}
+  \frac{\alpha ^2 r^2}{(r-1)^2}~\]
is in the {\rm limit circle} case near $r=1$ and hence is not essentially
 self-adjoint.
\end{lemma}
{\noindent {\it Proof.--}} Let $U$ be  the unitary transform
$U: u\ra r^{1/2} u $ from
$L^2(]0,1[,rdr )$ onto $L^2(]0,1[,dr )$.
Then $K= U HU^{-1}$ is given by
\[ -\frac{d^2}{dr^2}-\frac{1}{4 r^2} +
 \frac{\alpha ^2 r^{2}}{(r-1)^2}~.\]
$K$ is known to be in the limit circle case at $r=1$
(Theorem X.10 in \cite{RS}.)
\hfill $\square$

\subsection{Singular points}

\subsubsection{Monopoles}
We will first discuss the case of monopoles in 
$\bR ^3$. Here $\Omega $ is $\bR ^3 \setminus 0$.

\begin{definition}
The {\rm monopole of degree $m$}, $m\in  \bZ \setminus 0$,
is the  magnetic field 
$B_m=(m/2)p^\star (\sigma  )$ where
$p:\bR ^3 \setminus 0 \rightarrow  \bS^2 $ is the radial projection
and $\sigma $ the area form on $\bS^2$.
In coordinates
\[ B_m=\frac{m}{2}\frac{xdy\wedge dz +ydz\wedge dx +zdx \wedge dy
}{\left(x^2+y^2+z^2
\right)^{3/2}}~.\]
\end{definition}
\begin{remark}
Let us note, for  comparisons with the case where
$\pa \Omega $ is of codimension 1,  that $|B_m|_{\rm sp} = \frac{|m|}{2}
r^{-2} $.

\end{remark}
The flux of $B_m$ through $\bS^2$ is equal to $2\pi m$. This is
a well-known {\it quantisation condition} which is needed in order to
build a quantum monopole.
In order to define the Schr\"odinger operator $H_m$, we first 
introduce an Hermitian  complex line bundle $L_m$ with an
Hermitian  connexion $\nabla_m $ on $\Omega $ with curvature
$B_m$. We first construct $L_m$ and $\nabla_m $ on $\bS^2$
and then take their pull-backs: $\nabla_m $ in a direction tangent
to a sphere is the same and  $\nabla_m $ vanishes on radial
directions.
We have, using spherical coordinates,
\[ H_m=-\frac{\pa ^2}{\pa r^2}-\frac{2}{r}\frac{\pa }{\pa r }
+ \frac{1}{r^2}K_m ~,\]
where $K_m$ is the angular  Schr\"odinger operator on $\bS^2$ (discussed 
for example in \cite{Torki}).
Let us denote by $\lambda _1^m $ the lowest eigenvalue
of $K_m $. The self-adjointness of $H_m$ depends of the value 
of $\lambda _1^m  $.
As a consequence of Weyl's theory for Sturm-Liouville equations,
$H_m$ is \ess ~ if and only if  $\lambda _1^m \geq 3/4 $. 
From \cite{KU1,KU2,Torki} (sketched in Section 
\ref{sec:K_m}), we know
that
$\lambda _1^m =|m|/2$ so that
\begin{theorem}
The Schr\"odinger operator  $H_m$ (monopole of degree $m$)
 is \ess~ if and only
if $|m|\geq 2$.
\end{theorem}
\subsubsection{The spectra of the
 operators $K_m$,  the ``spherical Landau levels''} \label{sec:K_m}
These spectra are computed in \cite{KU1,KU2}
and in  the PhD thesis  \cite{Torki}.
We sketch here the calculus.
Recall that $K_m$ is the Schr\"odinger operator with magnetic
field $m \sigma /2 $ where $\sigma $ is the area form on $\bS^2$.
The metric is the usual Riemannian metric on $\bS^2$:
\begin{theorem}
The spectrum of $K_m$ is the sequence
\[ \lambda _k =\frac{1}{4}\left( k(k+2)-m^2 \right) ,~k=|m|, |m|+2,\cdots
~,\]
with multiplicities $k+1$.
In particular, the ground state $\lambda _{|m|}$
 of $K_m$ is $|m|/2$,
with multiplicity $|m|+1$. The ground state is exactly
the norm of the magnetic field. 

\end{theorem}
If $m=0$, the reader will recognise the spectrum of the Laplace
operator on $\bS^2$. 

We start with the sphere $\bS^3$ with the canonical metric.
Looking at $\bS^3\subset \bC^2$, we get an free isometric action
of $\bS^1_\theta $ on $\bS^3$: $\theta. (z_1,z_2)=  e^{i\theta }(z_1,z_2)$.
The quotient manifold is $\bS^2$ with $1/4$ times the canonical metric;
the volume $2\pi ^2$  of $\bS^3$ divided by $2\pi$ is $\pi $ which
is one  forth of $4\pi$. 

The quotient map $\bS^3 \rightarrow \bS^2$ is the Hopf fibration, 
 a $\bS^1-$principal
bundle. The sections of $L_m $ over $\bS^2$ are identified
with the functions on $\bS^3$ which satisfy
$f(\theta z)=e^{im\theta }f(z)$.
With this identification of the sections of $L_m$, we have
\[ K_m=\frac{1}{4} \left(  \Delta _{\bS^3}-m^2 \right) ~,\]
where $1/4$ comes from the fact that the quotient metric is
$1/4$ of the canonical one and $m^2$ from the action of
$\pa _\theta ^2$ which has to be removed. 
It is enough then to look at the spectral decomposition
of $\Delta _{\bS^3}$ using spherical harmonics:
 the $k$th eigenspace of $ \Delta _{\bS^3}$
is of dimension  $(k+1)^2 $ and splits  into
$k+1$ subspaces of dimension $k+1$ corresponding
to $m=-k, -k+2,\cdots , k$.

\subsubsection{A general result for $\Omega =\bR ^d \setminus 0$}

In this section $\Omega =\bR ^d \setminus 0$
and $B$ is singular at the origin.
\begin{theorem}\label{theo:singular}
If $\lim _{x\ra 0}|x|^{2}|B(x)|_{\rm sp}=+\infty $
 and, for any $x\ne 0$, 
 the direction $n(tx )$ has a limit
as $t\ra 0^+  $, then $M_B$ is \ess ~  
\end{theorem}
{\noindent {\it Proof.--}}
The proof is essentially the same as the proof of Theorem \ref{essd}
except that in the application of IMS method, we have to take a conical
partition of unity whose gradients can only be bounded by $|x|^{-1}$.
\hfill $\square$

\subsubsection{Multipoles} \label{sec:multi}

Let us denote, for $x\in \bR^3$, 
${\cal B}_x$ the monopole with centre $x$: ${\cal B}_x =\tau_x^\star
( B_2)$
with $\tau_x$ the translation by $x$ and ${ B}_2$ the  monopole 
with $m=2$.
If $P\left(\frac{\pa}{\pa x}\right)$
 is a homogeneous linear differential operator
 of degree $n$ on $\bR^3$ with constant
coefficients, we define
$ B_P=P ({\cal B}_x )_{x=0}$. 
Then $B_P $ is called a multipole of degree $n$.
All multipoles are exact!
It is a consequence of the famous Cartan's formula: if $P$ is of
degree 1, hence a constant vector field, 
\[B_V= {\cal L}_V{\cal B}_0 = d \left( \iota (V){\cal B}_0 \right) ~.\]
 A multipole of degree $1$ is called
a dipole;  viewed from very far away, the magnetic field of the earth
looks like a dipole.
 
\begin{theorem}
If $B_V=dA_V $ is a dipole ,
$H_{A_V}$ is   \ess.
\end{theorem}
{\noindent {\it Proof.--}}
Because $B_V$ is homogeneous of degree $-\alpha =-3$,
 it is enough, using \ref{theo:singular},  to show
that
$B_V$ does not vanish.
$V$ is a constant vector field, hence up to a dilatation, we can take
 $V=\pa / \pa z$.
We have 
\[ B_{\pa / \pa z}=\frac{d}{dt}_{|t=0}\frac{xdy\wedge dz+y dz\wedge dx+
(z-t)dx\wedge dy}{\left( x^2+y^2+(z-t)^2\right)^{3/2}} ~,\]
 which gives           
\[ B_{\pa / \pa z} =\frac{ 3xz dy\wedge dz + 3yz dz\wedge dx       
 + (2z^2 -x^2-y^2) dx \wedge dy }{(x^2+y^2+z^2)^{5/2}}~.\]
The form $B_{\pa / \pa z} $ does not vanish in $\Omega $.  
\hfill $\square$
\begin{remark}
We do not know if  all multipoles of degree $\geq 2 $  are  \ess . 
\end{remark}

\end{document}

%% file: etoile.pstex_t
\begin{picture}(0,0)%
\includegraphics{etoile.pstex}%
\end{picture}%
\setlength{\unitlength}{4144sp}%
\begingroup\makeatletter\ifx\SetFigFont\undefined%
\gdef\SetFigFont#1#2#3#4#5{%
  \reset@font\fontsize{#1}{#2pt}%
  \fontfamily{#3}\fontseries{#4}\fontshape{#5}%
  \selectfont}%
\fi\endgroup%
\begin{picture}(4360,2083)(1288,-1905)
\put(2048,-1088){\makebox(0,0)[lb]{\smash{{\SetFigFont{12}{14.4}{\rmdefault}{\mddefault}{\updefault}{\color[rgb]{0,0,0}$\pa \Omega $}%
}}}}
\put(3782,-585){\makebox(0,0)[lb]{\smash{{\SetFigFont{12}{14.4}{\rmdefault}{\mddefault}{\updefault}{\color[rgb]{0,0,0}$\pi $}%
}}}}
\put(4711,-765){\makebox(0,0)[lb]{\smash{{\SetFigFont{12}{14.4}{\rmdefault}{\mddefault}{\updefault}{\color[rgb]{0,0,0}$\Omega _\infty $}%
}}}}
\put(1461, 19){\makebox(0,0)[lb]{\smash{{\SetFigFont{12}{14.4}{\rmdefault}{\mddefault}{\updefault}{\color[rgb]{0,0,0}$e_2$}%
}}}}
\put(3206,-741){\makebox(0,0)[lb]{\smash{{\SetFigFont{12}{14.4}{\rmdefault}{\mddefault}{\updefault}{\color[rgb]{0,0,0}$e_n$}%
}}}}
\put(3216,-976){\makebox(0,0)[lb]{\smash{{\SetFigFont{12}{14.4}{\rmdefault}{\mddefault}{\updefault}{\color[rgb]{0,0,0}$e_0$}%
}}}}
\put(1336,-1836){\makebox(0,0)[lb]{\smash{{\SetFigFont{12}{14.4}{\rmdefault}{\mddefault}{\updefault}{\color[rgb]{0,0,0}$e_1$}%
}}}}
\end{picture}%

%% file: functionF.pstex_t
\begin{picture}(0,0)%
\includegraphics{functionF.pstex}%
\end{picture}%
\setlength{\unitlength}{4144sp}%
\begingroup\makeatletter\ifx\SetFigFont\undefined%
\gdef\SetFigFont#1#2#3#4#5{%
  \reset@font\fontsize{#1}{#2pt}%
  \fontfamily{#3}\fontseries{#4}\fontshape{#5}%
  \selectfont}%
\fi\endgroup%
\begin{picture}(5525,1421)(248,-905)
\put(616,-811){\makebox(0,0)[lb]{\smash{{\SetFigFont{12}{14.4}{\rmdefault}{\mddefault}{\updefault}{\color[rgb]{0,0,0}$\rho $}%
}}}}
\put(4350,-841){\makebox(0,0)[lb]{\smash{{\SetFigFont{12}{14.4}{\rmdefault}{\mddefault}{\updefault}{\color[rgb]{0,0,0}$R$}%
}}}}
\put(5123,-825){\makebox(0,0)[lb]{\smash{{\SetFigFont{12}{14.4}{\rmdefault}{\mddefault}{\updefault}{\color[rgb]{0,0,0}$R+1$}%
}}}}
\put(784,-806){\makebox(0,0)[lb]{\smash{{\SetFigFont{12}{14.4}{\rmdefault}{\mddefault}{\updefault}{\color[rgb]{0,0,0}$2\rho $}%
}}}}
\put(1202,-825){\makebox(0,0)[lb]{\smash{{\SetFigFont{12}{14.4}{\rmdefault}{\mddefault}{\updefault}{\color[rgb]{0,0,0}$1$}%
}}}}
\put(443,357){\makebox(0,0)[lb]{\smash{{\SetFigFont{12}{14.4}{\rmdefault}{\mddefault}{\updefault}{\color[rgb]{0,0,0}$F(u)$}%
}}}}
\put(5632,-617){\makebox(0,0)[lb]{\smash{{\SetFigFont{12}{14.4}{\rmdefault}{\mddefault}{\updefault}{\color[rgb]{0,0,0}$u$}%
}}}}
\put(293,172){\makebox(0,0)[lb]{\smash{{\SetFigFont{12}{14.4}{\rmdefault}{\mddefault}{\updefault}{\color[rgb]{0,0,0}$1$}%
}}}}
\put(263,-752){\makebox(0,0)[lb]{\smash{{\SetFigFont{12}{14.4}{\rmdefault}{\mddefault}{\updefault}{\color[rgb]{0,0,0}$0$}%
}}}}
\end{picture}%